\documentclass[12pt]{article}
\usepackage{graphics}
\usepackage{xcolor}
\usepackage{amssymb}
\usepackage{epsfig,graphicx}
\usepackage{color}
\usepackage{amsmath}
\usepackage{amsthm}
\usepackage{amsopn}
\usepackage{subfigure}
\def\uno{\mbox{1 \kern-.59em {\rm l}}}

\def\be{\begin{equation}}
	\def\ee{\end{equation}}
\def\bea{\begin{eqnarray}}
	\def\eea{\end{eqnarray}}

\begin{document}
\begin{center}
{\bf{\large  Quantum coherence and Leggett-Garg inequality
}}
\vskip 4em {\bf A. Jalal}\footnote{ jalal.ali68@gmail.com}, {\bf S. M. Fazeli}\footnote{mahdi.fazeli@gmail.com, smfazeli@qom.ac.ir}, and {\bf M. M. Ettefaghi}\footnote{mettefaghi@qom.ac.ir}
 \vskip 1em
 Department of Physics, University of Qom, Qom 371614-6611,
Iran

 \end{center}
 
  \vspace*{1.9cm}
\begin{abstract}
 In this paper, we attempt to establish the relationship between quantum coherence and the violation of the Leggett-Garg inequality. In particular, employing the Lindblad equation, we obtain the pseudo-density matrix for a damping system to study the effect of environment interaction on the violation of this inequality in a two-state quantum system. It is shown that the violation of the Leggett-Garg inequality can be observed as long as temporal evolution does not induce decoherence. This statement is independent of the initial state of the system. Furthermore,  similar to the Horodecki criterion for the CHSH inequality (R. Horodecki et al. Phys. Lett. {\bf A200}, 340), we study necessary and sufficient conditions for violating the Leggett-Garg inequality. Hereby, under the circumstance that the inequality violation occurs, an upper bound for the time interval between consecutive measurements with respect to the time scale of interaction with the environment (the relaxation time) is obtained.

\end{abstract}
\newpage
\newpage
\section{Introduction}
The development of modern quantum mechanics has had profound implications, engaging fields ranging from philosophy to physics and eliciting objections since its inception. Einstein, a pioneer of quantum theory, famously referred to the potential effects between two spatially separated particles as "spooky action at a distance." This concern led to the formulation of the Einstein-Podolsky-Rosen (EPR) paradox in 1935, a collaboration between Einstein, Podolsky, and Rosen \cite{r1}. The EPR paradox demonstrated that the principles of quantum mechanics could conflict with everyday observations and the classical laws that physicists had traditionally used to describe physical systems. It suggested that these surprising predictions might indicate the incompleteness of the standard quantum theory. 

The EPR paradox rests on two key assumptions: realism and locality, both of which are intuitively acceptable within classical physics. In 1964, building on the EPR assumptions, Bell introduced an inequality that provided a means to experimentally test these assumptions \cite{r2}. Today, experimental results based on Bell's theorem have demonstrated that the simultaneous assumptions of realism and locality can be violated in laboratory settings. These surprising outcomes are closely related to spatial correlations in quantum mechanics, particularly the phenomenon of entanglement. The relationship between the violation of Bell's inequality and entanglement is bidirectional: on the one hand, observing a violation of Bell's inequality requires the presence of entangled states; on the other hand, when a two-particle system violates Bell's inequality, it indicates the existence of entanglement between the particles. Thus, Bell's inequality serves as a qualitative criterion for identifying entanglement. 

The experimental violation of Bell's inequality conclusively demonstrates that the assumptions of realism and locality cannot simultaneously hold true at the microscopic scale. Paradoxes such as Schrödinger's cat and the measurement problem, as well as the absence of quantum phenomena at the macroscopic scale, have led some physicists to propose that realism may be different at the microscopic and macroscopic scales. To explore this idea, Leggett and Garg proposed a temporal version of Bell's inequality in 1985 to investigate realism at the macroscopic scale for single-particle systems \cite{r3}. In this context, the assumption of locality is replaced by the assumption of non-invasive measurement, which posits that the system's state remains unchanged as a result of measurement. While this assumption might seem natural at the macroscopic level, it does not align with quantum mechanics. 
 
Therefore, based on the validity of macroscopic realism and non-invasive measurement, an inequality known as Leggett-Garg inequality (LGI) arises from the temporal correlations of measurements taken at different times. Quantum systems that do not conform to these assumptions can violate this inequality \cite{r4}. 

A central question, similar to that posed by Bell's inequality, is: What leads to the violation of LGI? In fact, to evaluate macroscopic realism, it is crucial to ensure that measurements are non-invasive. For many macroscopic systems, such as SQUIDs in condensed matter physics, measurements tend to be invasive. However, employing a measurement technique based on the "ideal negative result" may fulfill the non-invasive measurement condition \cite{r5,r5.1}. In this method, the measurement apparatus is configured so that if a dichotomic observable, Q, for instance, has the eigenvalue +1, the system interacts with the apparatus. Otherwise, the system leaves the apparatus unaffected. Naturally, the viability of this "ideal negative result" method is not independent of the validity of the first assumption, namely, the existence of macroscopic realism.

Another approach to challenge realism using the LGI involves experiments with an ensemble of particles prepared in the same initial state and undergoing stationary temporal evolution \cite{r6,r7,r8,r9}. Stationary evolution implies that the evolved state depends solely on the time interval between the measurements. Also to address the challenge of non-invasive measurement, several experiments have been designed using indirect or weak measurement techniques \cite{r10,r11,r12}. 

Also LGI can serve as a criterion for detecting quantum coherence in the temporal evolution of a system \cite{r8,r16,r17,r18,r19}. Quantum coherence is a crucial quantum resource, underpinning many phenomena in quantum optics, quantum information, biological systems, and so on \cite{r13}. Initially, quantum coherence was quantified within the framework of quantum resource theory in reference \cite{r15}, and since then, various quantities have been introduced. Many of them are reviewed in Ref. \cite{r13}. Additionally, in Ref. \cite{r20}, an experimental protocol independent of the LGI has been proposed for detecting coherence in a superconducting qubit (SQUID).
 
Given the inevitability of environmental interactions affecting quantum systems in real-world conditions, in this paper, we explore the impact of decoherence induced by environmental interactions on the violation of a LGI. Specifically, we consider an ensemble of quantum systems interacting with an environment, where their temporal evolution is stationary. After the initial measurement, where the system is measured along a different direction than the operator $Q$, the system's state vector becomes a coherent superposition of the operator's eigenstates. If decoherence is absent during time evolution, carefully choosing measurement directions at subsequent times can reveal a violation of LGI. However, if the interval between successive measurements exceeds the timescale of environmental interactions, this violation is not expected to be observable. 

Therefore, we consider an ensemble of open quantum systems governed by the Lindblad master equation \cite{r21}. Using the pseudo-density matrix formalism \cite{r22}, we calculate the correlations between measurement results at different times. To investigate the violation of the ‌LGI, we consider an experiment involving four successive measurements. Here, temporal correlations replace spatial correlations, making the LGI analogous to the CHSH inequality \cite{r4}.

The Lindblad master equation has been previously employed in the study by Varma et al. \cite{VarmaEtAlLGI2020}, who investigated the theoretical upper bound of the LGI violation within post-selected, non-unitary dynamics. Additionally, the impact of dephasing on LGI violations in qubit systems has been briefly considered \cite{Wilde2012}. While the association between LGI violations and quantum coherence (specifically the violation of probability sum rules) is a recognized concept \cite{Halliwell2016}, the primary objective of this work is to transition from a qualitative understanding to a quantitative mapping of how environmental interactions suppress the violation of the LGI. Unlike previous studies, we provide a generalized framework using the complete Lindblad equation and the pseudo-density matrix (PDM) formalism. The novelty of the present study lies in three main aspects. First, we perform a systematic analysis of LGI violation under realistic open system conditions, explicitly incorporating environment-induced decoherence through the complete Lindblad equation for a two-level quantum system. Second, we establish a direct and quantitative relationship between quantum coherence and LGI violation. Third, by employing the PDM formalism, we derive a necessary and sufficient condition ($\mathcal{M}_{R}$) for LGI violation, analogous to the Horodecki criterion for the CHSH inequality \cite{r23}. This approach yields an explicit upper bound for the time interval ($\tau$) between successive measurements, functionally establishing a bridge between the resource theory of quantum coherence and the operational bounds of macroscopic realism.

In the following section, we provide an overview of LGI. In section \ref{s3} the Lindblad equation is introduced for the time evolution of open quantum systems. In section \ref{s4} the pseudo-density matrix formalism is developed. In section \ref{s5} the impact of environmental interactions on the violation of LGI is examined. Finally, in the last section we summarize our conclusions.

\section{LGI}\label{s2}

To derive the ‌LGI, we consider a two-state system where the observable \( Q \) can take values of \( \pm1 \). It's not necessary for the system to have only the state \( Q=\pm1 \); it suffices for \( \vert Q \vert \leq1 \) \cite{r6}. The two-time correlation function  $C_{ij}=\langle Q_{i}Q_{j}\rangle$ , where $ Q_{i(j)}=Q(t_{i(j)})$, is defined as follows: 
\begin{equation}\label{1}
 	C_{ij}=\sum_{Q_{i},Q_{j}=\pm1}^{ } Q_{i}Q_{j}P_{ij}(Q_{i},Q_{j}) 
 \end{equation}
 
Here, $P_{ij}(Q_{i},Q_{j})$ represents the joint probability of obtaining \( Q_i \) and \( Q_j \) as measurement outcomes at times \( t_i \) and \( t_j \), respectively. This definition is classical for \( C_{ij} \), and its quantum analogue can be defined as follows: 
 \begin{equation}\label{2}
 	C_{ij}=\frac{1}{2}\bigg\langle\bigg\lbrace Q_{i},Q_{j}\bigg\rbrace \bigg\rangle
 \end{equation}
  
Where  
$\left\lbrace Q_{i},Q_{j}\right\rbrace=Q_{i}Q_{j}+Q_{j}Q_{i}\ $

Now, to derive LGI, we consider the following two fundamental assumptions: 
\begin{itemize}
	\item Macrorealism per se: ``a macroscopic object which has available to it two or more macroscopically distinct states is at ‘almost all’ times in one of these states (the ‘almost all’ is necessary because one has to allow, in such a theory, for transits between the states; however, this complication can be taken into account in the analysis)"\cite{r5.1}
	\item Non-invasive Measurability: It is possible in principle to determine which of these states the system is in without any effect on the state itself or on the subsequent system dynamics.
\end{itemize}
  
In addition, there is another assumption inherent in the derivation of the Bell and Leggett-Garg theorems, which is entirely natural and its violation is far-fetched: the measurement outcome at time \( t \) is unaffected by experiments to be conducted in the future. In other words, causality operates forward in time, implying that past events can influence future events, but not vice versa. 

The assumption of non-invasive measurement is an idealization that is never achieved in practice. However, if we consider an experiment conducted with a statistical ensemble of systems, and at time \(t_1\), the measurement apparatus is designed such that a system with an eigenvalue of \(+1\) for the observable \(Q\) interacts with the device—while systems with eigenvalue \(-1\) do not interact—then the state of systems with eigenvalue \(-1\) can be determined without disturbance. Although this assumption remains ideal, if it holds approximately, it allows for determining the system's state without causing interference. By identifying the system's state in a similar manner at time \(t_2\), the temporal correlation \(C_{12}\) can be calculated based on the aforementioned conditions.
  
This type of measurement, known as non-invasive measurement, may be feasible on a macroscopic scale with a realistic assumption. For states that possess eigenvalues contrary to the measurement device's, their states can be determined without measurement, based on realism. Now, given that for an observable $Q$ only the values$\pm 1$ are possible, it can be easily found that, for example, for measurements at four separate times, the following inequality must hold:

\begin{equation}\label{3}
 	K_{4} \equiv C_{21}+C_{32}+C_{43}-C_{41}\leq 2
\end{equation}

In general, for measurements at n distinct times, it can be demonstrated \cite{r4}:
\begin{eqnarray}\label{4}
 	-n\leq K_{n}\leq n-2,
 \end{eqnarray}
for odd values of $n$ and \(n\geq3 \) and
 \begin{eqnarray}\label{5}
 	-(n-2)\leq K_{n}\leq n-2,
 \end{eqnarray}
for even values of $n$ and \( n\geq4 \), where \( K_n \) is defined as follows:

 \begin{equation}\label{6}
 	K_{n} \equiv C_{21}+C_{32}+C_{43}+\cdots+C_{n(n-1)}-C_{n1}.
 \end{equation}

In cases where the number of measurements is odd, only the upper bound of the inequality is significant, and the lower bound does not provide information because it is not violated even at the quantum level.

\section {Open Quantum System}\label{s3}

In the conventional formulation of quantum mechanics, the time evolution of a system is described by a unitary operator. This implies that there is no collapse, and phase coherence remains preserved throughout the process. However, any real physical system is not entirely isolated and interacts with its environment. Interaction with external degrees of freedom leads to energy dissipation, changes, and randomization of the phase. These effects can explain the absence of quantum properties at the macroscopic scale. In such cases, we are no longer working exclusively with pure states, and thus cannot operate within the Schrödinger picture using state vectors. Instead, we must employ the density matrix, which can represent a classical mixture of quantum states. It is clear that in such cases, the evolution will not be unitary.

The concepts of reduced density matrix and partial trace are crucial in studying open quantum systems. The total system and its environment constitute a closed system. To compute the environment-induced evolution for the desired system, we perform a partial trace over the environment's degrees of freedom in the total density matrix, reducing it to the reduced density matrix, which represents the system alone. 

We consider a system whose Hilbert space, Hamiltonian and density matrix are denoted by \( \mathcal{H} \), \( H \), and \( \rho \), respectively. In quantum mechanics, the general state of a system is described by the following density matrix:
\begin{equation}\label{7}
	\rho=\sum_{i}^{ } p_{i} |\psi_{i}\rangle \langle\psi_{i}|.
\end{equation}
Here, \( p_i \) represents the probability of the system being in the pure state $|\psi_{i}\rangle$. These probabilities \( p_i \) form a completely classical distribution. Furthermore, we represent the Hilbert space, Hamiltonian, and density matrix of the environment by  $\mathcal{H}_{E},\rho_{E}$ and $H_{E}$, respectively. For the total system and environment, we use $\mathcal{H}_{T},\rho_{T}$ and $H_{T}$, respectively.
 Figure 1 provides a schematic representation of the system and its environment.
\begin{figure}[!h]
	\centering 
	\includegraphics[scale=0.4]{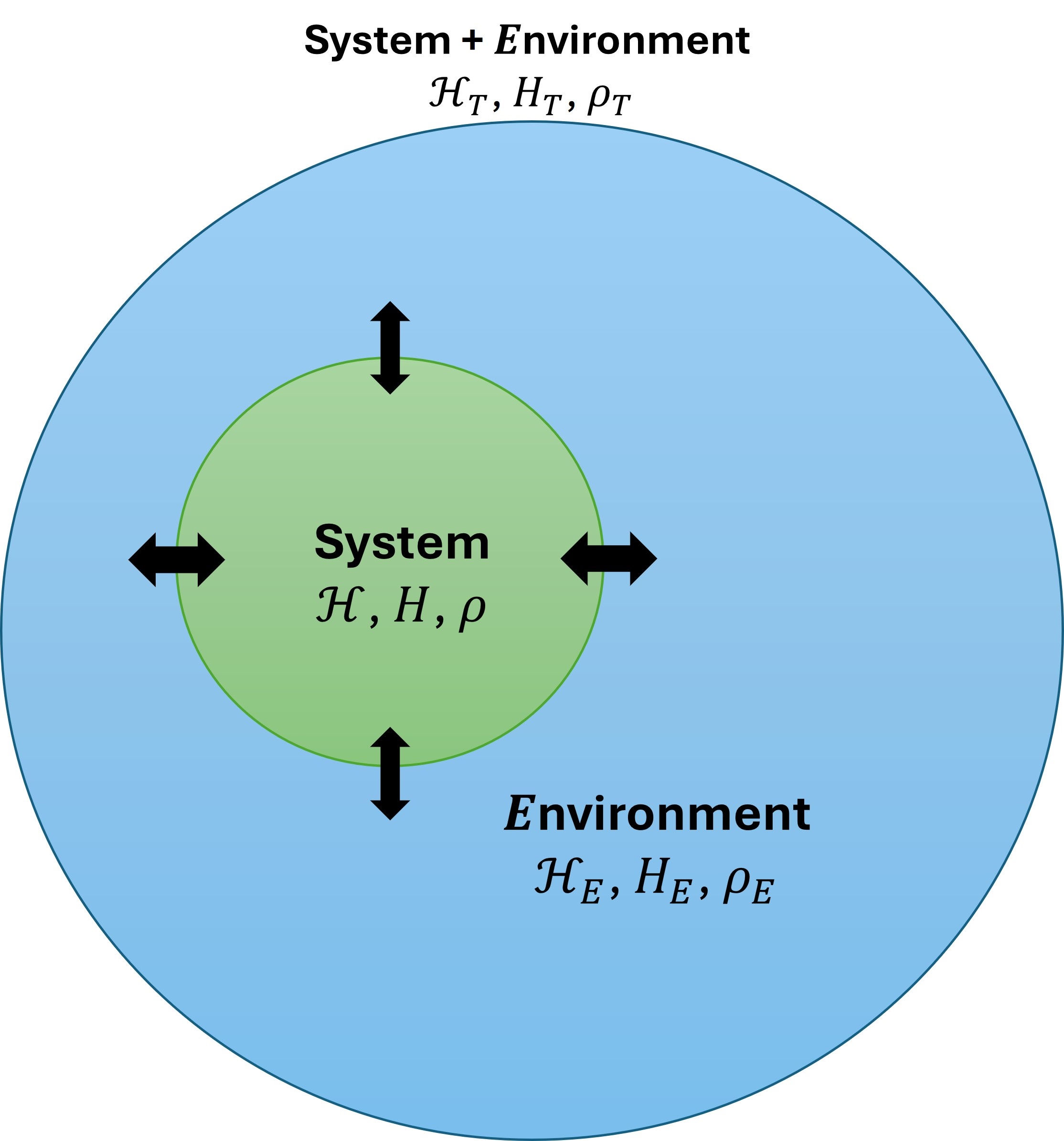}
	\caption{\small{A schematic representation of the system and its environment in open quantum system.}}
	\label{fig1}
\end{figure}

Several mathematical approaches have been developed to address the interaction between a quantum system and its surrounding environment. One of these methods is to consider the surrounding environment as a bath and the Markovian interaction of the system with the environment, which is generally expressed in the form of the Lindblad equation\cite{r21,r24}. In this formulation, the combined system and environment form a closed system. Consequently, the Hamiltonian of the entire system and environment can be expressed as:
\begin{equation}\label{8}
	H_{T}=H\otimes I+I\otimes H_{E}+H_{I}
\end{equation}
Here, \( H_{I} \) denotes the Hamiltonian describing the interaction between the system and the environment. Therefore, the time evolution of the system and environment is based on the Hamiltonian given in Eq. (\ref{8}). Partial tracing with respect to environment reveals that the evolution of the reduced density matrix for an N-dimensional system is governed by the following master equation:
 \begin{equation}\label{9}
 	\dot{\rho}=-\frac{i}{\hbar}[H,\rho]+\sum_{i=1}^{N^{2}-1} \gamma_{i} \left(\mathit{L}_{i}\rho\mathit{L}_{i}^{\dagger}-\frac{1}{2} \left\lbrace \mathit{L}_{i}^{\dagger}\mathit{L}_{i},\rho\right\rbrace  \right).
 \end{equation}

This equation is known as the Lindblad equation, where \( \mathit{L} \) represents the Lindblad superoperator, typically consisting of a set of jump operators that describe the dissipative dynamics. The form of the jump operators defines how the environment influences the system, and these operators are ultimately derived from microscopic models of system-environment interactions. The symbol $\left\lbrace~~\right\rbrace  $ denotes the anti-commutator, defined such that $\left\lbrace A,B\right\rbrace  =AB+BA$. Moreover, $\gamma_{i}$ represents a set of non-negative coefficients known as the damping  rates. According to Eq. (\ref{9}), the Lindblad equation reduces to the Von Neumann equation in the limit $\gamma_{i}\rightarrow 0$. For simplicity, \( \hbar \) is assumed to be equal to 1 throughout the calculations.

For a two-state system in the dipole  approximation $H_{I}=-\vec{D}.\vec{E}$, where \( \vec{D} \) represents the dipole operator of the system under investigation and \( \vec{E} \) is the electric field arising from the environment, the Lindblad equation is written as follows (for detailed derivations and considerations, see \cite{r21}):
\begin{eqnarray}\label{10}
	\dfrac{d}{dt}\rho(t) &=& -i[H,\rho(t)]+\gamma_{1}\bigg(\sigma_{-}\rho(t)\sigma_{+}-\frac{1}{2}\sigma_{+}\sigma_{-}\rho(t)-\frac{1}{2}\rho(t)\sigma_{+}\sigma_{-}\bigg)\notag\\
	& +&\gamma_{2}\bigg(\sigma_{+}\rho(t)\sigma_{-}-\frac{1}{2}\sigma_{-}\sigma_{+}\rho(t)-\frac{1}{2}\rho(t)\sigma_{-}\sigma_{+}\bigg)	
\end{eqnarray}

Here, $\sigma_{+}$ and $\sigma_{-}$ are the raising and lowering operators, respectively, which increase and decrease the system's energy by $\hbar\omega$ units. Their matrix representations are given by:

\begin{equation}\label{11}
	\sigma_{+(-)} = 
	\begin{pmatrix}
		0 & 1(0)  \\
		0(1) & 0	
	\end{pmatrix}.
\end{equation}

The precise calculation of  $\gamma_{i}$ can be found in reference \cite{r21}, but for our purposes in this work, precise calculation is not necessary. Instead, we introduce the values of these two parameters phenomenologically.

\section{Pseudo-Density Matrix}\label{s4}
 
To calculate the temporal correlation function, we require something akin to the density matrix for two different time instances. It's evident that the state of a system at two distinct times is generally different from a whole system composed of two subsystems—one corresponding to the first time and the other to the second time. Therefore, there's a need to introduce a new quantity that describes the state of the overall two-time system. The use of the pseudo-density matrix ($PDM$) formalism is essential for mapping the structure of temporal correlations into a unified matrix representation.This framework allows us to invoke eigenvalue analysis to derive the measurement-independent $\mathcal{M}_{R}$ criterion, a task that cannot be achieved through the direct calculation of individual correlation functions alone. For this purpose, we can utilize the formalism of the pseudo-density matrix introduced for instance in Ref. \cite{r22}. This quantity shares most properties of a density matrix, but its eigenvalues are not necessarily positive, hence termed a pseudo-density matrix. Now, let's briefly introduce the pseudo-density matrix.

The n-qubit matrices which are the $2^n\times2^n$ dimensional matrices composed of the direct product of the $2\times2$ dimensional Pauli and identity matrices, form a complete basis. Therefore, the density matrix of an n-qubit state can generally be written as:
\begin{equation}\label{12}
	\rho=a_{0} +\sum_{i}^{4^{n}-1 }a_{i} P_{i},
\end{equation}
where $P_{i}$'s, known as Pauli matrices in n-qubit space, are traceless matrices within this space. To normalize the density matrix, we must take $a_{0}=(\frac{\mathbb{I}}{2})^{n}$ in which $\mathbb{I}$ is $2\times2$ dimensional identity matrix. Given that the expectation value of an n-qubit Pauli matrix can be written as:
\begin{equation}\label{13}
\langle P_{j}\rangle=\text{Tr}\left( P_{j} \left( a_{0}+\sum_{i}^{4^{n}-1 }a_{i} P_{i}\right) \right)=2^{n} a_{j},
\end{equation}
the density matrix in Equation \eqref{12} can be expressed as:
\begin{equation}\label{14}
	\rho=\frac{\langle \mathbb{I}\rangle}{2^{n}}\mathbb{I}+ \sum_{i=1}^{4^{n}-1}\frac{\langle P_{i}\rangle}{2^{n}}P_{i}
\end{equation}
Therefore, based on the definition of $P_{i}$, the density matrix can be written as the direct product of single-qubit Pauli and identity matrices:
\begin{equation}\label{15}
	\rho=\frac{1}{2^{n}}\sum_{i_{1}=0}^{3}\cdots\sum_{i_{n}=0}^{3}\left\langle \prod_{j=1}^{n}\otimes\sigma_{i_{j}}\right\rangle   (\prod_{j=1}^{n}\otimes\sigma_{i_{j}}).
\end{equation}
Here, \(\sigma_0\) is the \(2\times2\) identity matrix, while the remaining matrices are the Pauli matrices. Eq. \eqref{15} can be regarded as the definition of the most general density matrix for an n-qubit system.

We can use the definition in \eqref{15} to introduce a more general density matrix: Let's imagine instead of dealing with an n-qubit system, we are working with a single-qubit system on which several Von Neumann measurements have been performed at various stages. Each Von Neumann measurement can be represented by single-qubit Pauli matrices. Consider a sequence of measurements $\left\lbrace \sigma_{i_{j}}\right\rbrace _{j=1}^{n}=\left\lbrace \sigma_{i_{1}}\cdots \sigma_{i_{n}}\right\rbrace$ performed on the system at different times. After the first measurement, the system collapses to one of the eigenstates $\sigma_{i_{1}}$ and then undergoes a time evolution until the second measurement. At this stage, the system again collapses to one of the eigenstates $\sigma_{i_{2}}$ and undergoes a time evolution until the next measurement. This process continues until the n-th measurement step.
So, if we replace the factor $\left\langle  \prod_{j=1}^{n}\otimes\sigma_{i_{j}} \right\rangle $ in Equation \eqref{15} with the expectation value of operators $\left\lbrace \sigma_{i_{j}}\right\rbrace _{j=1}^{n}$ calculated based on the described evolution, and replace $\left(\prod_{j=1}^{n}\otimes\sigma_{i_{j}}\right)$ with the product of measured operators, we get:
\begin{equation}\label{16} 
R=\frac{1}{2^{n}}\sum_{i_{1}=0}^{3}\cdots\sum_{i_{n}=0}^{3}\left\langle  \left\lbrace \sigma_{i_{j}}\right\rbrace_{j=1}^{n}\right\rangle \left(\prod_{j=1}^{n}\otimes\sigma_{i_{j}}\right).
\end{equation}
This matrix possesses all the properties of a density matrix with the exception of being positive definite. Therefore, it is referred to as a pseudo-density matrix, and it can be used to calculate the temporal correlations of measurements at different moments. If, instead of \( n \) measurements on a system at \( n \) different times, we perform these measurements at a single moment on an \( n \)-qubit system whose subsystems are non-interacting, the pseudo-density matrix \( R \) will become the same as the ordinary density matrix.

For instance, consider a single-qubit system where measurement $\sigma_{i_{1}}$  is performed at time \( t_1 \) and followed by another measurement $\sigma_{i_{2}}$  at time \( t_2 \). Let's assume the initial state is denoted by \( |\psi \rangle \). The probability that the measurement $\sigma_{i_{1}}$ at \( t_1 \) yields the eigenvalue \( +1 \) is $|\langle +_{\sigma_{i_{1}}}|\psi\rangle|^{2}$, and the eigenvalue \( -1 \) is $|\langle -_{\sigma_{i_{1}}}|\psi\rangle|^{2}$. If, as a result of the measurement, the eigenvalue $+1(-1)$ emerges, the initial state \( |\psi \rangle \) collapses to the eigenstate $|+_{\sigma_{i_{1}}} \rangle (|-_{\sigma_{i_{1}}} \rangle )$. Then it undergoes time evolution until \( t_2 \) when the operator $\sigma_{i_{2}}$ is measured. Hence, we have:
\begin{eqnarray}\label{17} 
\langle\left\lbrace \sigma_{k_{l}}\right\rbrace _{l=1}^{2} \rangle &=&(+1)|\langle +_{\sigma_{i_{1}}}|\psi\rangle|^{2}\langle +_{\sigma_{i_{1}}}|U^{\dagger}(t_{1},t_{2})\sigma_{i_{2}}U(t_{1},t_{2})|+_{\sigma_{i_{1}}}\rangle\notag\\
&+&(-1)|\langle -_{\sigma_{i_{1}}}|\psi\rangle|^{2}\langle-_{\sigma_{i_{1}}}|U^{\dagger}(t_{1},t_{2})\sigma_{i_{2}}U(t_{1},t_{2})|-_{\sigma_{i_{1}}}\rangle.
\end{eqnarray}
It is evident that if no measurement is performed initially, Eq. (\eqref{17}) is simplified to:
\begin{equation}\label{18} 
\langle\left\lbrace \sigma_{k_{l}}\right\rbrace _{l=1}^{2} \rangle =\langle \psi|U^{\dagger}(t_{1},t_{2})\sigma_{i_{2}}U(t_{1},t_{2})|\psi\rangle,
\end{equation}
and in the absence of a measurement at the second step, it becomes:
\begin{equation}\label{19} 
\langle\left\lbrace \sigma_{k_{l}}\right\rbrace _{l=1}^{2} \rangle =(+1)|\langle +_{\sigma_{i_{1}}}|\psi\rangle|^{2}+(-1)|\langle -_{\sigma_{i_{1}}}|\psi\rangle|^{2}.
\end{equation} 

These equations correspond to a situation where the time evolution is unitary. However, if the system’s time evolution is non-unitary, as occurring for an open quantum systems, we must incorporate the effects of this evolution through the density matrix formalism. Explicitly, let's assume the initial state of the system is described by the density matrix \(\rho(t_{1}; t_{1})\) and the operator \( \sigma_{i_{1}} \) is measured on this system at time \( t_1 \). Consequently, during this measurement, the system's state collapses to one of the eigenstates of this operator. The density matrix corresponding to the resulting states is denoted by \(\rho_{+_{i_{1}}}(t_{1};t_{1})\) and \(\rho_{-_{i_{1}}}(t_{1};t_{1})\). These states evolve according to the Lindblad equation, and at time \( t_2 \), the second measurement, i.e., the operator \( \sigma_{i_{2}} \), is performed on the system. In this case, the coefficients of the pseudo-density matrix are determined as follows:
\begin{eqnarray}\label{20} 
\langle\left\lbrace \sigma_{k_{l}}\right\rbrace _{l=1}^{2} \rangle &=&\text{Tr}\bigg(\rho(t_1,t_1)~\sigma_{i}~\rho_{+_{i}}(t_{1};t_{1})\bigg)~\text{Tr}\bigg(\rho_{+_{i}}(t_{2};t_{1})~\sigma_{j}\bigg)\notag\\
&+&\text{Tr}\bigg(\rho(t_1,t_1)~\sigma_{i}~\rho_{-_{i}}(t_{1};t_{1})\bigg)~\text{Tr}\bigg(\rho_{-_{i}}(t_{2};t_{1})~\sigma_{j}\bigg).\notag\\
\end{eqnarray}
It's clear that if no measurement occurs at the second step, the second factor on the right-hand side of the equation above becomes equal to one. However, if no measurement is made at the first step, we can extend Equation \eqref{18} as follows:
\begin{eqnarray}\label{21} 
\langle\left\lbrace \sigma_{k_{l}}\right\rbrace _{l=1}^{2} \rangle &=&\text{Tr}\bigg(\rho(t_{2};t_{1})~\sigma_{i_{2}}\bigg).
\end{eqnarray}
Thus, obtaining the coefficients in Eq. \eqref{16}, one can construct the pseudo-density matrix between times \( t_1 \) and \( t_2 \).

Using the pseudo-density matrix, temporal correlations like Equation \eqref{fig2} can be expressed as follows:
\begin{equation}\label{22} 
C_{ij}=\text{Tr}\bigg(R_{ij} (Q_{i} \otimes Q_{j}) \bigg) 
\end{equation}
Here, \( R_{ij} \) represents the pseudo-density matrix at the initial time \( t_i \) and the subsequent time \( t_j \).

\section{Violation of the ‌LGI in an open quantum system}\label{s5}

Consider a two-state system whose time evolution is governed by the following Hamiltonian:
\begin{equation}\label{23} 
\hat{H}=\frac{1}{2}\omega\hat{\sigma}_{x}.
\end{equation}

For simplicity, hereafter we take \(\omega = 1\) in the following discussion. We assume that all measurements are performed along the z-axis direction of the spin component operator. Consequently, all \( Q_i \)'s are equal to \( \sigma_3 \). Furthermore, we assume that the time interval between all successive measurements is the same and denoted by \( \tau \). In other words, during the \( n \) measurement stages, for the \( i \)th and \( j \)th measurements ($j>i$), we have:
\begin{equation}\label{24} 
t_{j}-t_{i}=(j-i)\tau.
\end{equation}
Thus, within the framework of quantum mechanics, where measurements are entirely invasive and the system is considered isolated, we can demonstrate that:
\begin{equation}\label{25} 
C_{ij}=\cos\left((j-i)\tau\right).
\end{equation}
\( K_n \) defined in equation \eqref{6}, comes as follows:
\begin{equation}\label{26} 
K_{n}=(n-1)\cos(\tau)-\cos(n-1)\tau.
\end{equation}

\begin{figure}[!h]
	\centering 
	\includegraphics[scale=1.2]{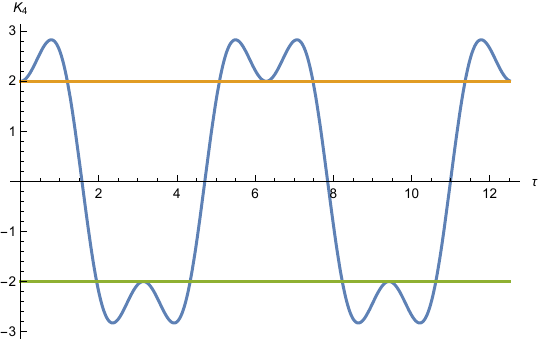}
	\caption{\small{Violation of the ‌LGI for a two-state system in the framework of isolated quantum mechanics. All measurements are performed along the z-direction. The system evolves with Hamiltonian $\hat{H}=\frac{1}{2}\hat{\sigma}_{x}$.}}
	\label{fig2}
\end{figure}

In Figure \ref{fig2}, is plotted \( K_4 (n=4) \). If our theory is consistent with the LGI postulates, according to Eq. \eqref{5}, we would expect that  \( |K_4| \leq 2\). However, in Figure \ref{fig2}, it is observed that for certain choices of \( \tau \), the value of \( K_4 \), computed based on isolated quantum mechanics, exceeds the mentioned limit.

As shown in Figure \ref{fig2}, for an isolated quantum system where decoherence is absent, the maximum value of the violation reaches $2\sqrt{2}$ and repeats periodically at intervals of $\tau = m\pi\pm\pi/4$. This periodic behavior arises because the nonzero Hamiltonian induces a temporal rotation of the system’s quantum state, causing the angle between the evolving state and the fixed measurement axis to periodically satisfy the condition for maximal violation. 
This phenomenon is analogous to the violation observed in the CHSH inequality, where appropriately chosen measurement settings lead to it. In particular, the violation with the maximal quantum bound occurs for a $\pi/4$ angle difference between the two measurement axes.  Hence, at a given moment, selecting the appropriate measurement axes allows us to observe the maximum violation. Meanwhile, in the case of LGI, the measurement direction can be assumed constant while the system rotates due to the time evolution.

As mentioned, real systems are not isolated and interact with their environment. Therefore, in what follows, we address the violation of the ‌LGI in an open quantum system. Accordingly, we assume that the time evolution of the considered system is given by the following Lindblad equation:
\begin{eqnarray}\label{27} 
\dfrac{d}{dt}\rho(t) &=& -i[H,\rho(t)]+\gamma_{1}\bigg(\sigma_{-}\rho(t)\sigma_{+}-\frac{1}{2}\sigma_{+}\sigma_{-}\rho(t)-\frac{1}{2}\rho(t)\sigma_{+}\sigma_{-}\bigg)\notag\\
& +&\gamma_{2}\bigg(\sigma_{+}\rho(t)\sigma_{-}-\frac{1}{2}\sigma_{-}\sigma_{+}\rho(t)-\frac{1}{2}\rho(t)\sigma_{-}\sigma_{+}\bigg)\notag\\
& +&\gamma_{3} \bigg(\sigma_{z}\rho(t)\sigma_{z}-\frac{1}{2}\sigma_{z}\sigma_{z}\rho(t)-\frac{1}{2}\rho(t)\sigma_{z}\sigma_{z}\bigg). 
\end{eqnarray}

Using this equation, the density matrix at different times can be calculated. If we have the density matrix at each moment, using Eq. (\ref{16}) we can construct the pseudo-density matrix between any two times \( t_i \) and \( t_j \), denoted as \( R_{ij} \). In this way, using Eq. \eqref{22}, we can calculate the correlation between the measurement results of the observable operators \( Q_i \) and \( Q_j \) performed at times \( t_i \) and \( t_j \), respectively.

\begin{figure}[!h] 
\centering 
\includegraphics[scale=1.2]{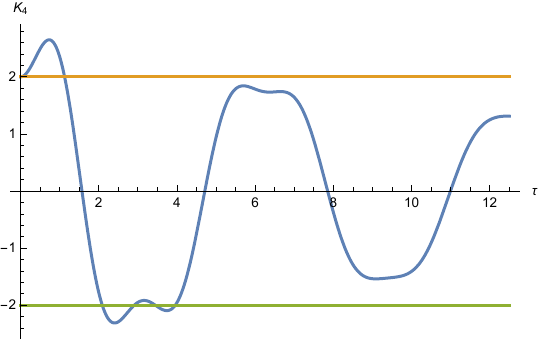}
\caption{\small{Violation of the LGI for a two-state system in the framework of open quantum mechanics (The unit of \(\tau\) is the inverse of the damping parameters \(\gamma_i\).). All measurements are performed along the z-direction, and $\gamma_{1} = \gamma_{2} = 2 \gamma_{3} = 0.03 $.}}
\label{fig3}
\end{figure}

We assume that the  initial state is completely incoherent
\begin{equation}\label{28}  
	\rho_{0}=\frac{1}{2}\begin{pmatrix}
		1 & 0 \\
		0 & 1 
	\end{pmatrix}.
\end{equation}
Moreover, similar to what was done for the closed quantum state, we assume the same time interval between each successive measurement ($t_{m+1}-t_{m}=\tau$). We have calculated the parameter \( K_4 \) and depicted it in terms of \( \tau \) through Figure \ref{fig3}. 
We see that although the violation of the ‌LGI is observed for small \( \tau \) due to the incoherence of the initial state, this violation disappears if \(\tau \) is chosen large enough. In other words, interaction with the environment leads to decoherence. Therefore, by examining the possibility of violating the LGI, we can obtain a criterion for decoherence. In fact, although unitary time evolution in quantum mechanics allows an initially incoherent state to develop coherence—manifested as nonzero off-diagonal elements in the density matrix—the onset of quantum decoherence gradually suppresses the periodic amplitude of the violation over time, eventually rendering quantum effects unobservable. This result illustrates the expected transition from quantum-coherent to classical behavior induced by environmental decoherence and is consistent with the discussions presented in Refs. \cite{r25,r26,r27}.

To investigate the possibility of violating the inequality governing \( K_4 \), we use a theorem given in reference \cite{r23}. Although this theorem is stated for the CHSH inequality, it is extendable to the ‌LGI governing \( K_4 \). Generally, the pseudo-density matrix of a two-time two-state system can be represented as a $4\times4$ matrix, analogous to a two-qubit system. Therefore, as a special case of Eq. \eqref{16}, it can be written based on the Pauli matrices of two qubits as follows:
\begin{equation} \label{29} 
R_{ij}=\frac{1}{4}\bigg(\mathbb{I}_{i}\otimes \mathbb{I}_{j}+\bf{r}.\bf{\sigma}_{i}\otimes \mathbb{I}_{j}+\mathbb{I}_{i}\otimes \bf{s}.\bf{\sigma}_{j}+\sum_{m,n=1}^{3}t_{mn}\sigma_{i_{m}}\otimes\sigma_{j_{n}}\bigg).
\end{equation}
Here, according to the notation adopted in Section 4, the identifiers \( i \) and \( j \) refer to the measurement stage. \( \vec{r} \) and \( \vec{s} \) vectors are defined in the $\mathbb{R}^{3}$ space. The \( t_{mn}=\text{Tr}(R_{ij}\sigma_{i_{m}}\otimes\sigma_{j_{n}}) \) coefficients form a real matrix, denoted as \( T_R \). We define the parameter $\mathcal{M}_R$ as follows:
\begin{equation} \label{30} 
\mathcal{M}_R=u_{k}+u_{l}
\end{equation}
Here, \( u_k \) and \( u_l \) represent two largest eigenvalues of matrix $U=T_{R}^{\dagger}T_{R}$. Similar to the approach outlined in reference \cite{r23}, it can be demonstrated that the maximum value of \( K_4 \) by changing (angle of) $Q_{i}$ operators, corresponds to \( 2\sqrt{\mathcal{M}_R} \). Hence, according to Eq. \eqref{5}, a violation of the ‌LGI is possible if \( \mathcal{M}_R > 1 \). 

\begin{figure}[!h] 
\centering 
\includegraphics[scale=1.2]{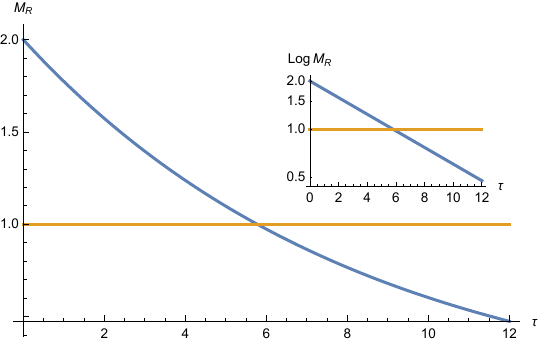}
\caption{\small{The quantity $\mathcal{M}_R$ as a function of $\tau$ for a two-state system in the framework of open quantum mechanics. $\gamma_{1} = \gamma_{2} = 2 \gamma_{3} = 0.03 $.}}
\label{fig4}
\end{figure}

Using the Lindblad time evolution equation (Eq. \eqref{27}), we have computed $\mathcal{M}_R$ with damping parameters \(\gamma_i\) similar to those chosen for Figure \ref{fig3} and depicted it with respect to \(\tau\) in Figure \ref{fig4}(Here, without loss of generality, we consider the Hamiltonian to be zero for simplicity. This is justified by considering the interaction picture \cite{r21}.).

It is observed that, given the damping parameters \(\gamma_i\), measuring the violation of the LGI becomes impossible after a certain time. The reason is straightforward: interaction with the environment causes the quantum coherence of the system’s state to vanish after passing time $\tau$ since the first measurement. Therefore, the environment decoherence leads to negate the conditions required for violating the LGI. Consequently, this inequality can serve as a qualitative criterion for assessing coherence preservation during the system’s evolution.

We note that although the amplitude of temporal correlation combinations such as $K_4$ is damped due to the interaction of the quantum system with its environment, this behavior differs from the quantum decay processes discussed in Ref. \cite{r28}. A detailed investigation of LGI violations in genuine quantum decay processes, including the effects of environmental decoherence, lies beyond the scope of the present work but represents an interesting direction for future studies.

\section{Possible Physical Implementations}

The theoretical framework developed in this work, based on the Lindblad
master equation and the pseudo-density-matrix formalism, can be directly
applied to several physical systems where decoherence and relaxation are
experimentally accessible. In particular, violations of the LGI under decoherence have been experimentally observed in a
variety of quantum platforms. For example, two-level Rydberg atoms coupled to high-$Q$
resonators have provided early evidence for LGI violation in open systems \cite{r7}, photonic systems demonstrate temporal quantum correlations through weak
measurement techniques \cite{r11}, and superconducting flux qubits exhibit obvious LGI violations even in the presence of environmental noise \cite{r20}.

From an experimental perspective, the theorem developed in this work provides a theoretical benchmark for assessing the maximum possible violation of the LGI in a given quantum system. In typical LGI tests, one must design an appropriate measurement configuration and optimize the measurement axes to achieve maximal violation. Our approach, however, determines this upper bound directly from the system’s pseudodensity matrix, without requiring prior specification of the measurement directions. Therefore, the theorem offers experimentalists a clear theoretical reference indicating whether LGI violation is possible in principle and how large it can be, guiding the design and interpretation of practical measurement setups.

Within the Lindblad framework adopted in this work, the damping parameters 
$\gamma_i$ characterize distinct environment-induced processes. Specifically, 
$\gamma_{1}$ and $\gamma_{2}$ describe population relaxation and excitation between 
the two energy levels, while $\gamma_{3}$ accounts for pure dephasing, which destroys 
the phase coherence without altering the populations. In conventional terminology, 
the total relaxation rate defined by $\Gamma_{1}=\gamma_{1}+\gamma_{2}$ and the overall 
decoherence rate given by $\Gamma_{2}=\Gamma_{1}/2+\gamma_{3}$ are experimentally accessible
(see, e.g., \cite{r21,r29}).

For a thermal environment (Thermal Bath) at temperature $T$, the relationship between the two population rates ($\gamma_1$ and $\gamma_2$) is determined by the Boltzmann distribution, which is necessary to ensure the system relaxes to a thermal equilibrium state:
\begin{equation}
	\label{eq:boltzmann}
	\frac{\gamma_2}{\gamma_1} = e^{-\frac{\hbar\omega}{k_B T}}.
\end{equation}
Alternatively, this ratio can be expressed in terms of the mean number of thermal quanta ($N_{th}$) in the bath, which is given by the Bose-Einstein distribution:
\begin{equation}
	N_{th} = \frac{1}{e^{\frac{\hbar\omega}{k_B T}} - 1},
\end{equation}
such that the rates $\gamma_1$ (decay rate from excited state) and $\gamma_2$ (excitation rate from ground state) are then related by:
\begin{equation}
	\frac{\gamma_2}{\gamma_1} = \frac{N_{th}}{N_{th} + 1}.
\end{equation}
One of these thermodynamic constraints provides the third essential equation to uniquely determine the three damping parameters ($\gamma_1, \gamma_2, \gamma_3$) from the two experimental relaxation and decoherence rate ($\Gamma_1, \Gamma_2$) and the temperature ($T$).
For many quantum experiments (e.g., superconducting qubits) where the temperature is very low ($k_B T \ll \hbar\omega$), the Boltzmann factor $e^{-\frac{\hbar\omega}{k_B T}} \approx 0$. In this low-temperature limit, $\gamma_2$ is negligible ($\gamma_2 \approx 0$), as thermal excitation is highly suppressed \cite{r21}.

 Open-system dynamics of the type studied here play an essential role in various biological and mesoscopic contexts, such as excitonic energy transfer in photosynthetic complexes \cite{r4,r30}. In such scenarios, the measurable decoherence rates correspond directly to the $\gamma_i$		
parameters in our framework, enabling a quantitative comparison between experimental observations and theoretical predictions.

\section{Conclusion}\label{s6}

The LGI provides a qualitative criterion for assessing the validity of macroscopic realism. Its violation suggests the presence of coherent superposition at the macroscopic scale. Given that quantum coherence is the resource for LGI violation, the inequality can be interpreted as a qualitative criterion for coherence. While LGI was designed for macroscopic systems, its investigation at the microscopic scale does not conflict with its underlying postulates, provided the non-invasive measurement assumption is substituted by the condition of stationarity for an ensemble of quantum systems.

In real-world experimental setups, the effect of decoherence, resulting from the interaction between the quantum system and its environment, must be accounted for. Consequently, this paper investigates the LGI violation in an open quantum system by employing the Lindblad equation for temporal evolution and constructing the pseudo-density matrix to calculate temporal correlations. While a closed system shows violation by choosing an appropriate time interval $\tau$ (see Figure 2), this is generally not the case for open systems; as Figure 3 shows, the violation diminishes if $\tau$ exceeds the timescale of environmental interaction.

To reach a general conclusion, we employed the method outlined in reference \cite{r23}  to analyze the maximum possible violation. The key contributions of this study are summarized as follows:

\begin{enumerate}
	\item We established a direct and quantitative relationship between quantum coherence and LGI violation in open systems.
	\item We utilized the pseudo-density matrix formalism to derive a necessary and sufficient criterion ($\mathcal{M}_{R}$) for LGI violation, which provides a robust, measurement setting-independent method analogous to the Horodecki criterion for entanglement.
	\item Our analysis demonstrated a fundamental constraint: for LGI violation to be observable, the time interval between consecutive measurements ($\tau$) must be smaller than the system's decoherence timescale ($\tau \le \tau_{decoherence}$).
\end{enumerate}

This result provides an explicit, practical upper bound for experimental design, functionally establishing the LGI as an operational criterion for identifying the quantum-to-classical transition governed by environmental interaction. Our results provide a direct theoretical framework for analyzing Leggett-Garg violations in realistic experimental systems such as superconducting qubits and photonic platforms, where decoherence parameters are measurable via the Lindblad formalism.

\end{document}